\documentclass[a4paper,11pt]{article}
\usepackage{pos}

\title{Hadronization studies at the LHC with ALICE }
\author[a, b]{Jianhui Zhu for the ALICE Collaboration}

\affiliation[a]{Institute of Particle Physics, Central China Normal University,\\
  NO.152 Luoyu Road, Wuhan, China}

\affiliation[b]{GSI Helmholtz Centre for Heavy Ion Research,\\
Planckstraße 1, Darmstadt, Germany}

\emailAdd{zjh@ccnu.edu.cn}

\abstract{Studies of the production of heavy-flavour baryons are of prominent importance to investigate hadronization mechanisms at the LHC, in particular through the study of the evolution of the baryon-over-meson production ratio. Measurements performed in pp and p--Pb collisions at the LHC have revealed unexpected features, qualitatively similar to what was observed in heavy-ion collisions and, in the charm sector, not in line with the expectations based on previous measurements from $\rm e^+e^-$ colliders and in ep collisions. These results suggest that charmed baryon formation might not be universal and that the baryon-over-meson ratio depends on the collision system or multiplicity.

A review of ALICE measurements of charmed baryons, including $\rm \Lambda_c^+/D^0$ as a function of charged-particle multiplicity in pp, p--Pb and Pb--Pb collisions, $\rm \Sigma_c^{0, +, ++}/D^0$ and $\rm \Xi_c^{0, +}/D^0$ as a function of $p_{\rm T}$ in pp collisions and $\rm \Gamma(\Xi_c^0\rightarrow\Xi^-e^+\nu_e)/\Gamma(\Xi_c^0\rightarrow\Xi^-\pi^+)$, will be presented. Comparison to phenomenological models will be also discussed. Emphasis will be given to the discussion of the impact of these studies on the understanding of hadronization processes.}

\FullConference{%
  40th International Conference on High Energy physics - ICHEP2020\\
  July 28 - August 6, 2020\\
  Prague, Czech Republic (virtual meeting)
}

\begin{document}
\maketitle

\section{Introduction}
The production of heavy-flavour hadrons in high-energy hadronic collisions can provide important tests of the theory of quantum chromodynamics (QCD) because perturbative techniques are applicable down to low transverse momentum ($p_{\rm T}$) due to the large mass of the heavy quark compared to the QCD scale parameter ($\rm \Lambda_{QCD}\sim200~MeV$). The production cross sections of heavy-flavour hadrons can be calculated using the factorisation approach as a convolution of three factors \cite{COLLINS198637}: the parton distribution functions (PDFs) of the incoming protons, the hard-scattering cross section at partonic level, calculated as a perturbative series in powers of the strong coupling constant $\alpha_{\rm s}$, and the fragmentation functions of heavy quarks into corresponding heavy-flavour hadrons, which is an inherently non-perturbative process related to, or even driven by, the confining property of QCD. The production cross sections of D- and B-meson in pp collisions can be described by several calculations adopting different factorisation schemes, such as the fixed order with next-to-leading-log resummation (FONLL) approaches \cite{Cacciari_1998, Cacciari_2012}, the general-mass variable flavour number scheme (GM-VFNS) \cite{Kramer:2017gct, Helenius:2018uul} and $k_{\rm T}$-factorisation \cite{CATANI1991135, _uszczak_2009, Maciu_a_2013} perturbative quantum chromodynamics (pQCD) calculations. However, due to the lack of knowledge about the fragmentation function of heavy quarks into baryonic states, some of these calculations do not provide predictions for heavy-baryon production. The heavy-flavour baryon-to-meson ratio is an ideal observable related to the hadronization mechanism since the contributions from parton distribution function and parton-parton scattering terms are cancelled in the ratio. The measurement of charmed mesons in pp collisions is described by calculations \cite{Acharya:2017jgo}, in which the fragmentation fractions are obtained from $\rm e^+e^-$ and ep collisions, indicating that charm fragmentation may be universal. However, the $\rm \Lambda_c^+/D^0$ ratio at the LHC is found to be enhanced with respect to predictions based on $\rm e^+e^-$ and $\rm ep$ experiments, suggesting that the charm fragmentation functions are not universal among different collision systems. A similar observation has been made in the measurement of the inclusive $\rm \Xi_c^0$ baryon production at mid-rapidity in pp collisions at $\sqrt{s}=7$~TeV \cite{Acharya:2017lwf}. Several hadronization mechanisms, such as colour reconnection (CR) beyond leading colour approximation \cite{Christiansen:2015yqa}, coalescence \cite{Plumari:2017ntm} and feed-down from a largely augmented set of higher mass charm-baryon states beyond the current listings of the particle data group (PDG) \cite{He:2019tik, He:2019vgs}, have been proposed to explain this enhancement. Newest measurements of charmed baryons $\rm \Lambda_c^+$, $\rm \Sigma_c^{0, ++}$, $\rm \Xi_c^{0, +}$ performed with the ALICE experiment will be used to verify these hadronization mechanisms.

\section{Experimental apparatus and data analysis}
The ALICE apparatus is equipped with various detectors for triggering, tracking and particle identification (PID) \cite{Aamodt:2008zz}. The main detectors used for these measurements are the Inner Tracking System (ITS), the Time Projection Chamber (TPC) and the Time-Of-Flight detector (TOF), which are located in the central barrel at mid-rapidity ($|\eta|<0.9$) embedded in a solenoidal magnet that provides a $\rm B=0.5$~T field parallel to the beam direction. The ITS is used for tracking, vertex reconstruction and particle identification (PID) via energy loss (${\rm d}E/{\rm d}x$) measurements. The TPC is the main tracking detector in the central barrel and is also used for PID. The TOF provides further complementary PID informations via measurements of the time of flight. The reconstruction of charmed baryons is performed with both hadronic and semi-leptonic decay channels. PID and decay topological selections with either conventional linear combination of selection criteria or machine learning techniques are implemented to reject background. The signal extraction in hadronic decay channels is performed by invariant mass analyses. The $\rm \Xi_c^0$ is also measured via its semi-leptonic decay channel ($\rm \Xi_c^0\rightarrow\Xi^-e^+\nu_e$) and the invariant mass distribution is obtained by subtracting wrong-sign pairs from right-sign pairs and Bayesian unfolding is used to correct the missing electron neutrino momentum. After signal extraction, the yields are corrected for acceptance and efficiency and are converted into cross sections. The non-prompt fraction is corrected for all charmed baryons except the $\rm \Xi_c^0$.

%\section{Charmed baryons signal extraction}
%\begin{figure}[!ht]
%\includegraphics[width=0.37\textwidth]{figures/2020-05-22-LcInvMass_46_HM.pdf}
%\includegraphics[width=0.4\textwidth]{figures/2020-06-15-mass_plot_Sc.pdf}
%\includegraphics[width=0.5\textwidth]{figures/2020-05-21-Xic0_InvMass_pp13TeV_pt3_4.png}
%\includegraphics[width=0.4\textwidth]{figures/2020-05-18-InvariantMass_XicPlus_0.pdf}
%\caption{The signal extraction of charmed baryons.}
%\label{fig:InvMass_Lc}
%\end{figure}

\section{Results}
\subsection{$\rm \Lambda_c^+$ production in pp, p--Pb and Pb--Pb collisions}

\begin{figure}[!ht]
\centering
\includegraphics[width=0.43\textwidth]{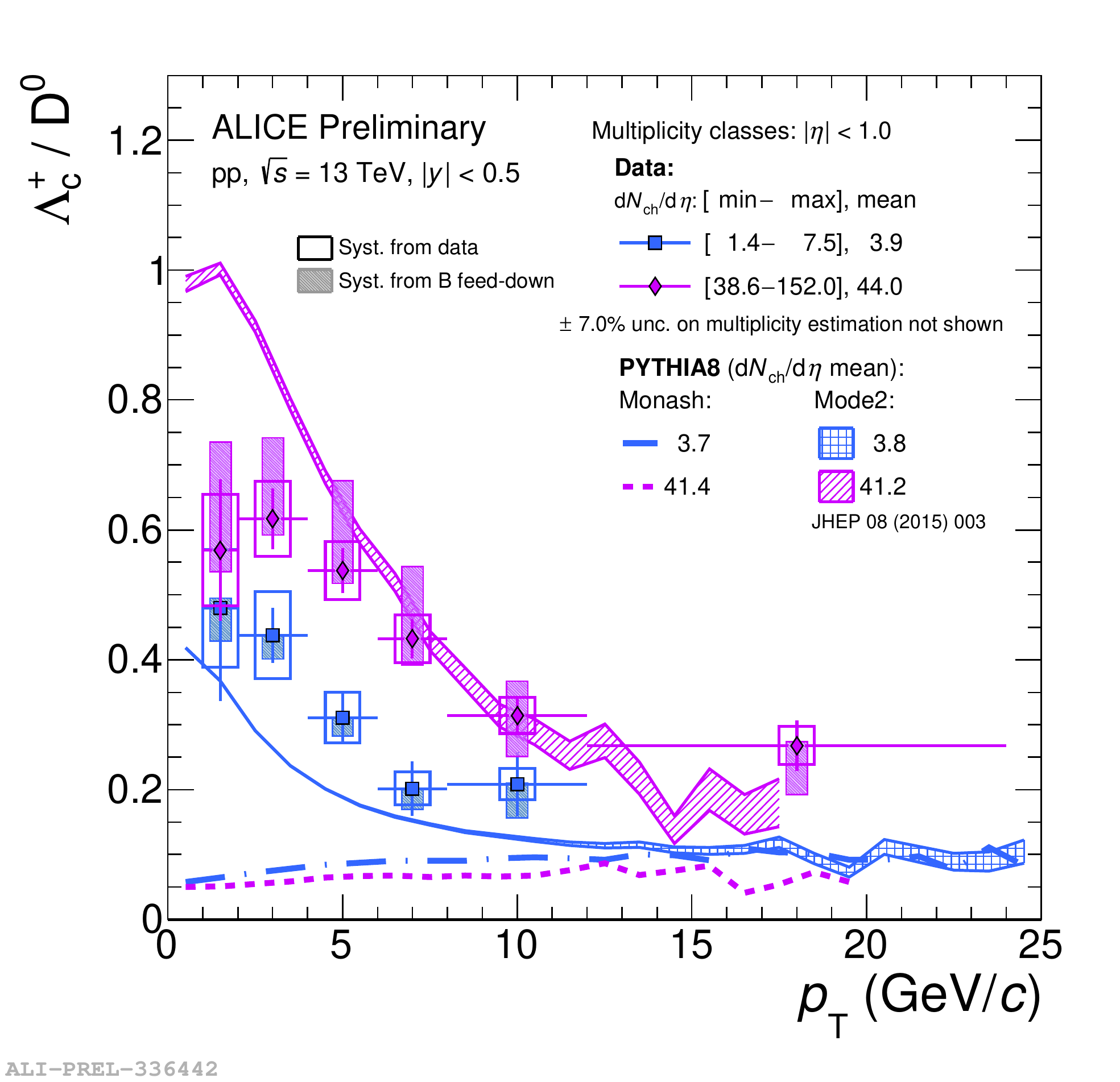}
\includegraphics[width=0.55\textwidth]{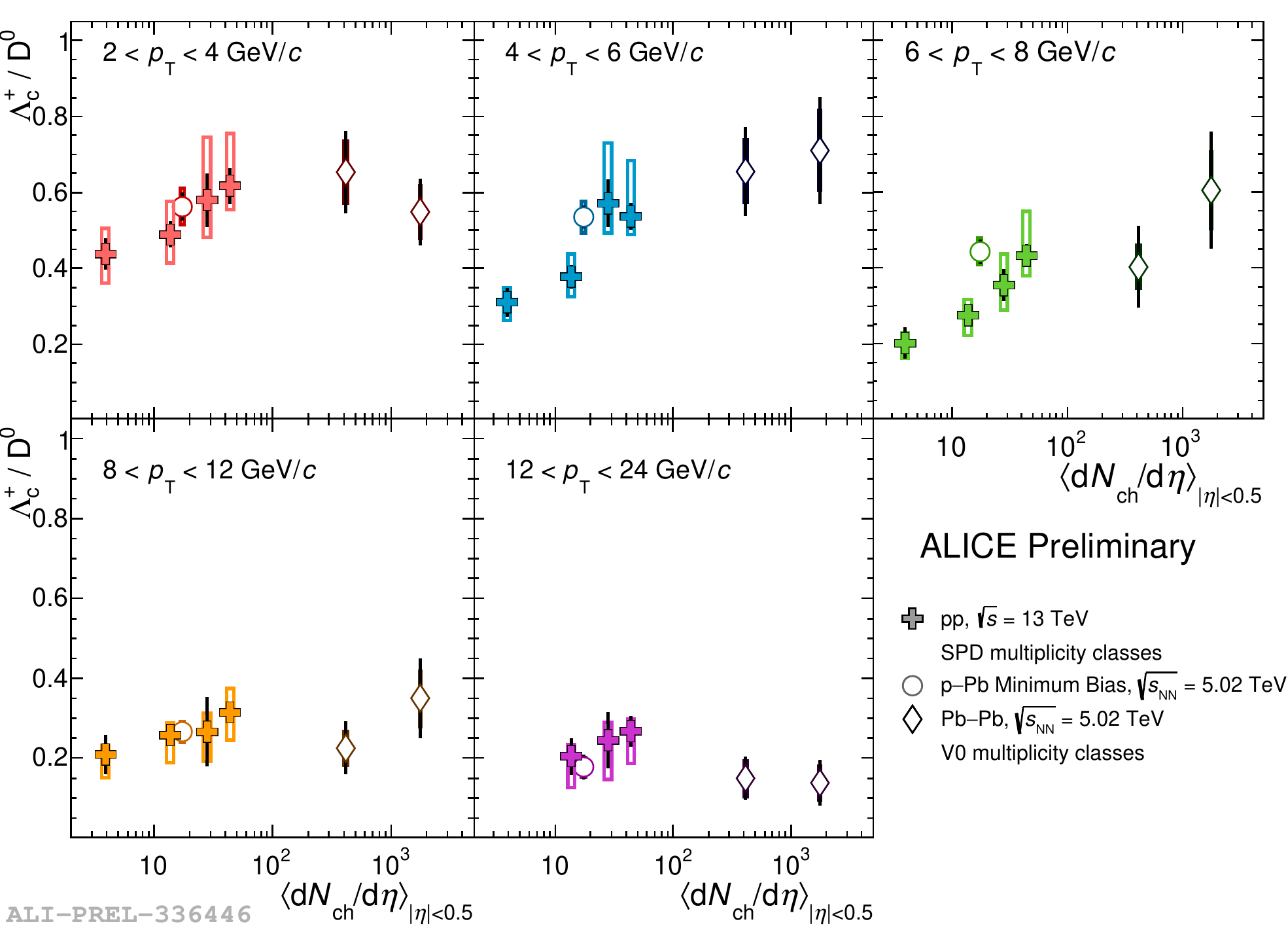}
\caption{(Left) $\rm \Lambda_c/D^0$ ratio measured in pp collisions at $\sqrt{s}=13$~TeV in different multiplicity intervals compared with their corresponding PYTHIA8 predictions \cite{Skands:2014pea, Christiansen:2015yqa}. (Right) $\rm \Lambda_c/D^0$ ratio measured in pp, p--Pb and Pb--Pb collisions as a function of the charged-particle multiplicity and for different $p_{\rm T}$ intervals.}
\label{fig:LcToD0}
\end{figure}

The $\rm \Lambda_c^+$ is reconstructed via the hadronic decay channels $\rm \Lambda_c^+ \rightarrow pK^-\pi^+$ (with branching ratio, $\rm BR=(6.28\pm0.32)\%$) and $\rm \Lambda_c^+ \rightarrow pK_s^0$ ($\rm BR=(1.59\pm0.08)\%$). The $\rm \Lambda_c/D^0$ ratio is measured as a function of $p_{\rm T}$ in different charged-particle multiplicity intervals in pp collisions at $\sqrt{s}=13$~TeV as shown in Fig.~\ref{fig:LcToD0} (left) and as a function of charged-particle multiplicity in pp collisions at $\sqrt{s}=13$~TeV, in p--Pb and Pb--Pb collisions at $\sqrt{s_{\rm NN}}=5.02$~TeV as shown in Fig.~\ref{fig:LcToD0} (right). In the left panel of Fig.~\ref{fig:LcToD0}, from low to high multiplicity, a significant enhancement is observed at low and intermediate $p_{\rm T}$ for the $\rm \Lambda_c/D^0$ ratio. This enhancement is not described by the standard PYTHIA8 Monash tune \cite{Skands:2014pea}, which is tuned based on the $\rm e^+e^-$ measurements. However, the PYTHIA8 Mode2 with colour reconnection using string formation beyond leading colour approximation \cite{Christiansen:2015yqa} provides a qualitative description of the data points. In the right panel of Fig.~\ref{fig:LcToD0}, the $\rm \Lambda_c/D^0$ ratio increases smoothly from low multiplicity pp to Pb--Pb collisions. However in high multiplicity pp collisions, this ratio is consistent with the one obtained in Pb--Pb collisions, indicating that the enhancement of $\rm \Lambda_c/D^0$ could depend on multiplicity rather than collision systems.

\subsection{$\rm \Sigma_c^{0, +, ++}$ production in pp collisions at $\sqrt{s}=13$~TeV}

\begin{figure}[!ht]
\centering
\includegraphics[width=0.49\textwidth]{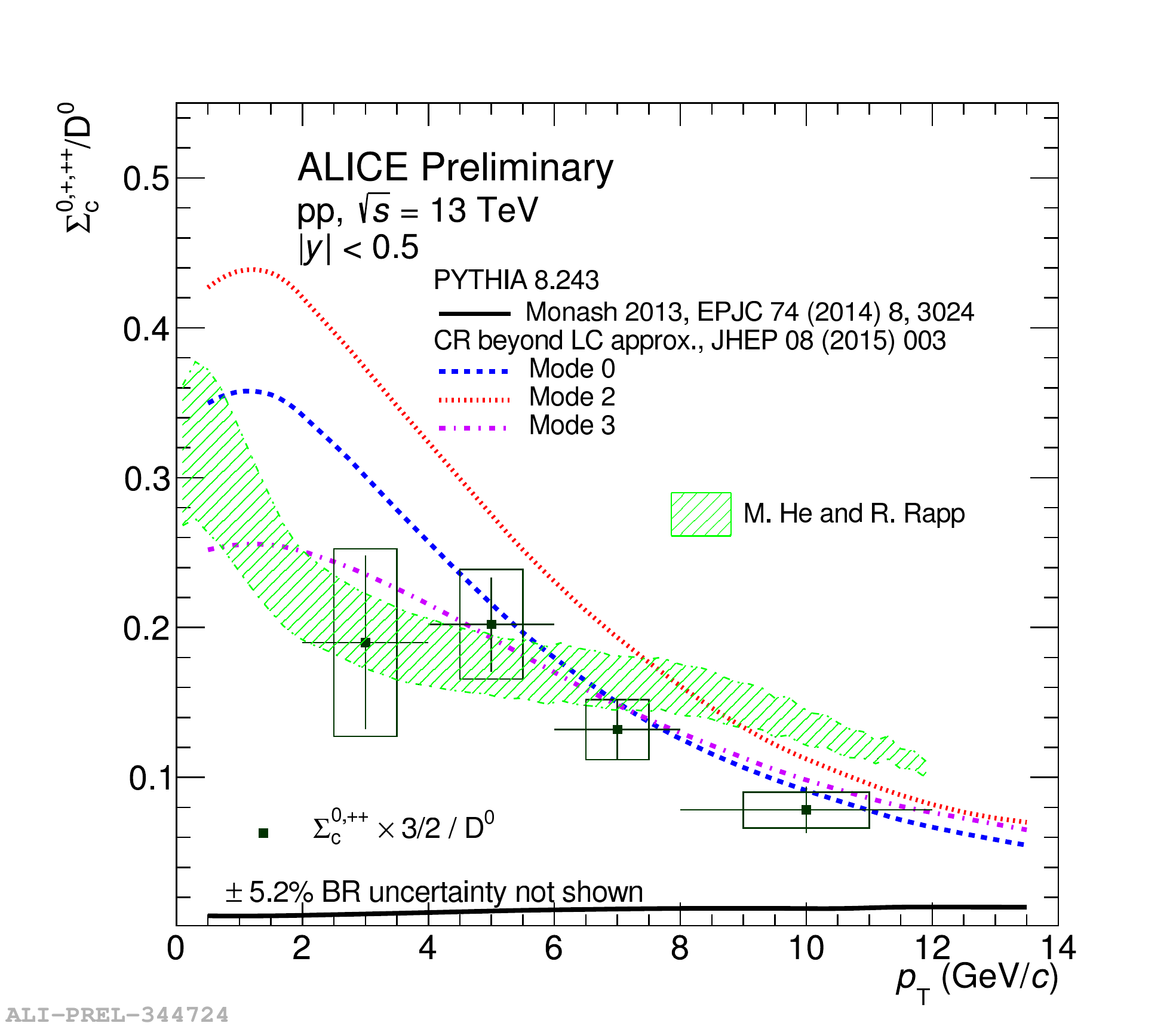}
\includegraphics[width=0.49\textwidth]{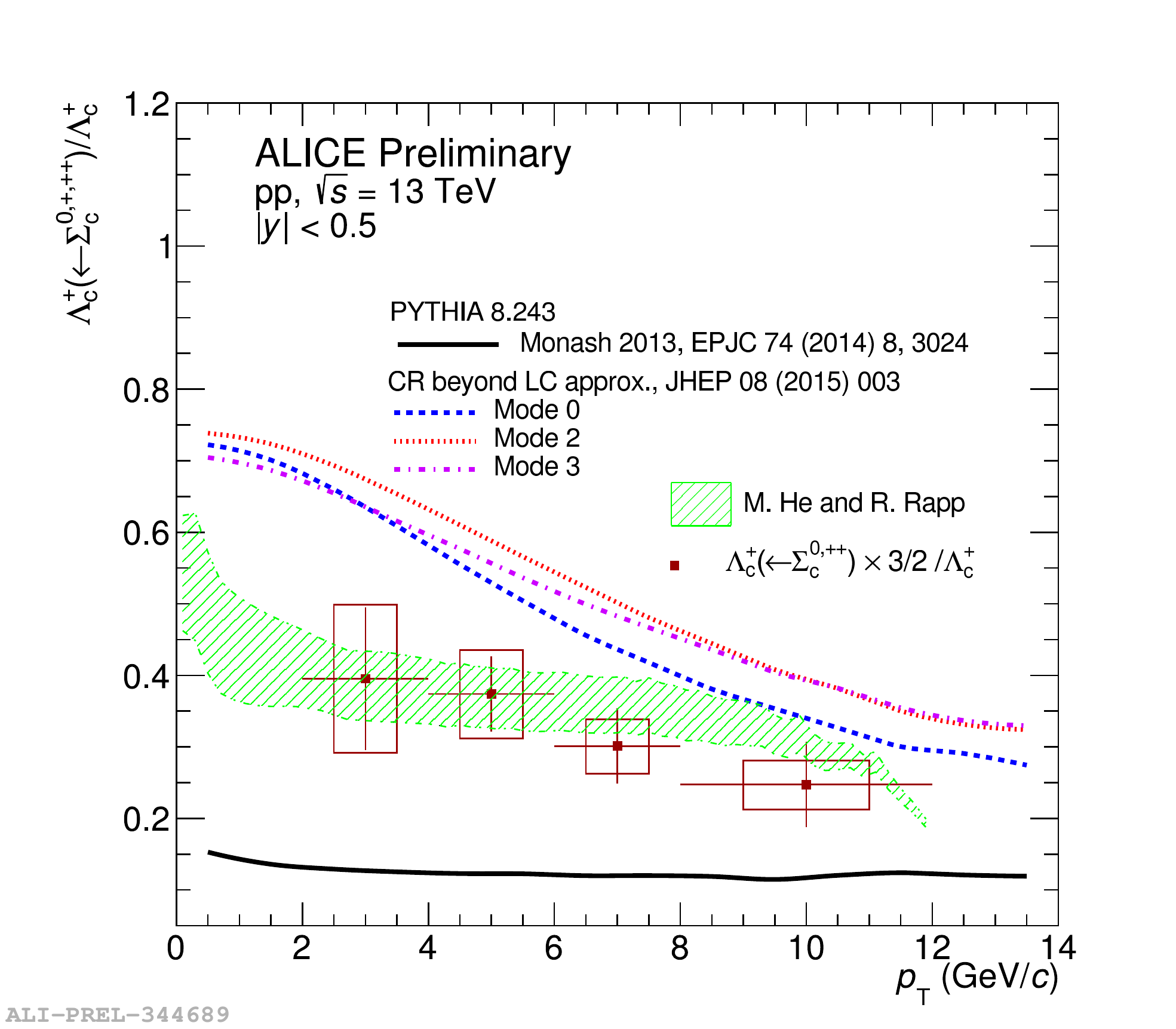}
\caption{$\rm \Sigma_c^{0, +, ++}/D^0$ ratio (left) and $\rm \Lambda_c^+(\leftarrow \Sigma_c^{0, +, ++})/\Lambda_c^+$ ratio (right) measured in pp collisions at $\sqrt{s}=13$~TeV compared with theoretical models.}
\label{fig:SigmacToD0}
\end{figure}

The $\rm \Sigma_c^{0, ++}$ baryons are reconstructed via the hadronic decay channels $\rm \Sigma_c^0 \rightarrow \Lambda_c^+\pi^-$ ($\rm Br\approx100\%$) and $\rm \Sigma_c^{++} \rightarrow \Lambda_c^+\pi^+$ ($\rm Br\approx100\%$), respectively. In order to compare with theoretical calculations conveniently, $\rm \Sigma_c^{0, +, ++}$ is calculated by multiplying $\rm \Sigma_c^{0, ++}$ by $3/2$ according to the isospin symmetry. The measurement of $\rm \Sigma_c^{0, +, ++}$ production is very important to constrain the production of $\rm \Lambda_c^+$ since $\rm \Sigma_c^{0, +, ++}$ almost completely decays to $\rm \Lambda_c^+$. In the left panel of Fig.~\ref{fig:SigmacToD0}, the $\rm \Sigma_c^{0, +, ++}/D^0$ ratio is measured as a function of $p_{\rm T}$ in pp collisions at $\sqrt{s}=13$~TeV compared with PYTHIA8 predictions and the statistical hadronization model (SHM) \cite{He:2019tik}, employing a largely augmented set of charm-baryon states beyond the current listings of the particle data group. The $\rm \Sigma_c^{0, +, ++}/D^0$ ratio shows a similar $p_{\rm T}$ trend as the previous $\rm \Lambda_c^+/D^0$ ratio. All three CR modes and the SHM describe the trend of data, but the standard Monash tune significantly underestimates the data by a factor $\sim$15-20 at low $p_{\rm T}$ and a factor $\sim$5 at high $p_{\rm T}$. In the right panel of Fig.~\ref{fig:SigmacToD0}, the fraction of $\rm \Lambda_c^+$ produced from $\rm \Sigma_c^{0, +, ++}$ decays ($\rm \Lambda_c^+(\leftarrow \Sigma_c^{0, +, ++})/\Lambda_c^+$) is measured as function of $p_{\rm T}$ in pp collisions at $\sqrt{s}=13$~TeV compared with PYTHIA8 and SHM predictions. All three CR modes overestimate this ratio and the Monash tune underestimates this ratio. However, the SHM is in good agreement with data. Fig.~\ref{fig:SigmacToD0} indicates that the enhancement of $\rm \Lambda_c^+/D^0$ ratio in pp collisions can be partially explained by $\rm \Sigma_c^{0, +, ++}$ feed-down. So far, the SHM works better than PYTHIA8 with colour reconnection in heavy flavour hadronization mechanism in pp collisions.

\subsection{$\rm \Xi_c^{0, +}$ production in pp collisions at $\sqrt{s}=13$~TeV}

The $\rm \Xi_c^0$ is reconstructed via both the semi-leptonic decay channel $\rm \Xi_c^0\rightarrow\Xi^-e^+\nu_e$ ($\rm Br=(1.8\pm1.2)\%$), which is analysed using similar techniques to that reported in \cite{Acharya:2017lwf}, and the hadronic decay channel $\rm \Xi_c^0\rightarrow\Xi^-\pi^+$ ($\rm Br=(1.43\pm0.32)\%$), which is reconstructed via an invariant mass analysis. The $\rm \Xi_c^+$ is reconstructed via the hadronic decay channel $\rm \Xi_c^+\rightarrow\Xi^-\pi^+\pi^+$ ($\rm Br=(2.86\pm1.21\pm0.38)\%$ \cite{Li:2019atu}). Fig.~\ref{fig:XicToD0} (left) shows the $\rm \Xi_c^0/D^0$ and $\rm \Xi_c^+/D^0$ ratios as a function of $p_{\rm T}$. They are consistent with each other. All theoretical models underestimate the measured $p_{\rm T}$-differential $\rm \Xi_c^{0, +}$ ratio. The ratios of $\rm \Lambda_c^+/D^0$ and $\rm \Sigma_c^{0, +, ++}/D^0$ obtained from the Monash tune significantly underestimate the data by a factor $\sim$20-30 in the low $p_{\rm T}$ region and by a factor $\sim$3-4 in the highest $p_{\rm T}$ interval. This new $\rm \Xi_c^{0, +}$ measurement provides important constraints to models of charm quark hadronisation in pp collisions.

\subsection{$\rm \Gamma(\Xi_c^0\rightarrow\Xi^-e^+\nu_e)/\Gamma(\Xi_c^0\rightarrow\Xi^-\pi^+)$ in pp collisions at $\sqrt{s}=13$~TeV}

\begin{figure}[!ht]
\centering
\includegraphics[width=0.49\textwidth]{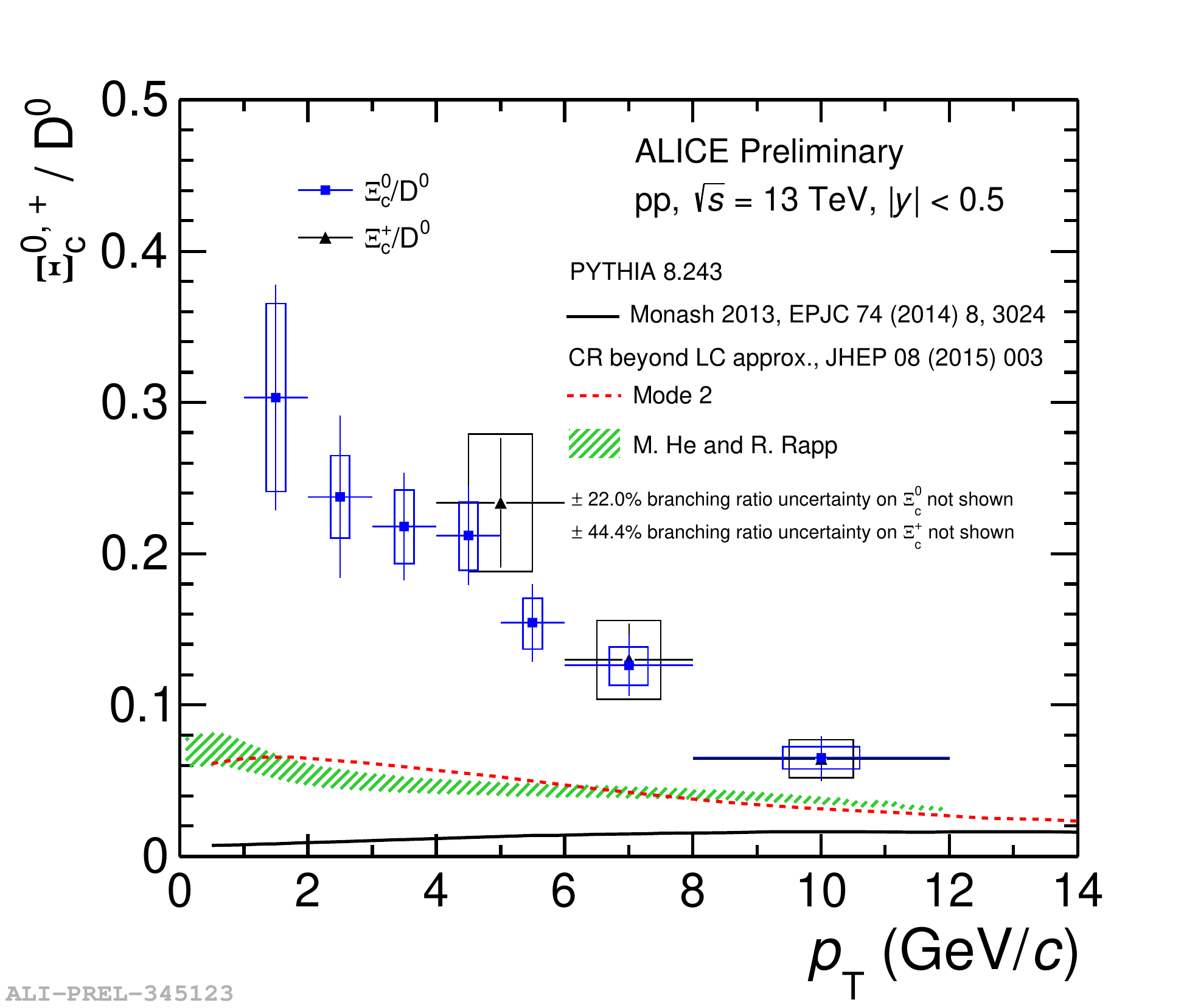}
\includegraphics[width=0.49\textwidth]{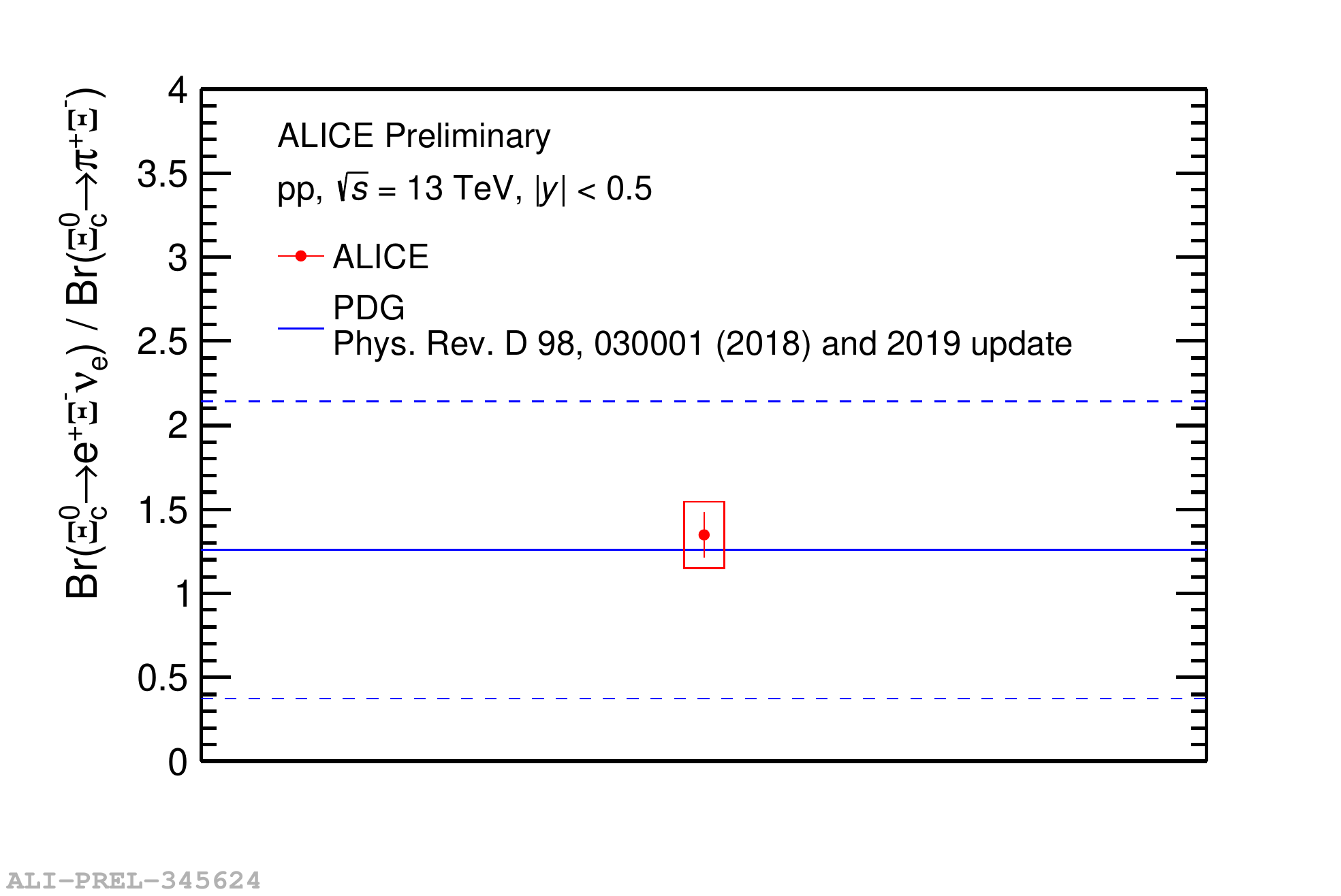}
\caption{(Left) $\rm \Xi_c^{0, +}/D^0$ ratio measured in pp collisions at $\sqrt{s}=13$~TeV compared with theoretical models. (Right) Ratio of semi-leptonic and hadronic decay branching ratios measured in pp collisions at $\sqrt{s}=13$~TeV compared with the PDG value.}
\label{fig:XicToD0}
\end{figure}

The $\rm \Gamma(\Xi_c^0\rightarrow\Xi^-e^+\nu_e)/\Gamma(\Xi_c^0\rightarrow\Xi^-\pi^+)$ is measured from the $p_{\rm T}$-dependent ratio between the two corrected yields. The total $p_{\rm T}$ correlated systematic uncertainty is computed shifting up and down the $p_{\rm T}$ dependent ratio of the yields for the $p_{\rm T}$ correlated uncertainty of the two measurements. The final systematical uncertainty of the ratio is obtained by summing in quadrature of $p_{\rm T}$ correlated and $p_{\rm T}$ uncorrelated systematical uncertainty. Fig.~\ref{fig:XicToD0} (right) shows the ratio measured by ALICE, which is consistent with the PDG value (1.3 $\pm$ 0.8) and demonstrates $\sim$4 times smaller uncertainty.

\subsection{Summary and outlook}
Charmed baryon-to-meson ratios $\rm \Lambda_c^+/D^0$, $\rm \Sigma_c^{0, +, ++}/D^0$ and $\rm \Xi_c^{0, +}/D^0$ in pp collisions at $\sqrt{s}=13$~TeV measured by ALICE have been presented. PYTHIA8 Monash tune is insufficient to describe the hadronization processes, however Mode tunes with higher-order colour reconnection perform better on this. Another model based on statistical hadronization with augmented charmed baryon states also provides qualitatively good predictions for $\rm \Lambda_c/D^0$ and $\rm \Sigma_c/D^0$. However, none of models can describe the enhancement of $\rm \Xi_c$ baryons with respect to $\rm D^0$. The $\rm \Gamma(\Xi_c^0\rightarrow\Xi^-e^+\nu_e)/\Gamma(\Xi_c^0\rightarrow\Xi^-\pi^+)$ measured by ALICE is consistent with the PDG value and has $\sim$4 times smaller uncertainty.

\section{Acknowledgments}
This work is supported by the international postdoctoral exchange fellowship program by the Office of China Postdoctoral Council (No. 20181016), the National Natural Science Foundation of China (Grant No. 11805079) and national key research and development program under Grant (No. 2018YFE0104700).

% Reduce space between lines in Reference %
\let\oldthebibliography\thebibliography
\let\endoldthebibliography\endthebibliography
\renewenvironment{thebibliography}[1]{
  \begin{oldthebibliography}{#1}
    \setlength{\itemsep}{0.03em}
    \setlength{\parskip}{0em}
}
{
  \end{oldthebibliography}
}

\bibliographystyle{utphys}
\bibliography{references}

\end{document}